\documentclass[twocolumn,showpacs,preprintnumbers,amsmath,amssymb]{revtex4}

\usepackage{graphicx}
\usepackage{dcolumn}
\usepackage{bm}

\begin{document}

\preprint{}

\title{A novel experimental method for the measurement of the caloric curves of clusters }

\author{Fabien Chirot$^{1,2}$}
\author{Pierre Feiden$^{1,2}$}
\author{S\'ebastien Zamith$^{1,2}$}
\author{Pierre Labastie$^{1,2}$}
\author{Jean-Marc L'Hermite$^{1,2}$}
\affiliation{%
$^1$ Universit\'e de Toulouse ; UPS ; 118 route de Narbonne, F-31062
Toulouse, France\\
$^2$ CNRS ; Laboratoire Collisions Agr\'egats R\'eactivit\'e, IRSAMC
; F-31062 Toulouse, France
}%

\date{\today}

\begin{abstract}
A novel experimental scheme has been developed in order to measure
the heat capacity of mass selected clusters. It is based on
controlled sticking of atoms on clusters. This allows one to
construct the caloric curve, thus determining the melting
temperature and the latent heat of fusion in the case of first-order
phase transitions. This method is model-free. It is transferable to
many systems since the energy is brought to clusters through
sticking collisions. As an example, it has been applied to
$Na_{90}^+$ and $Na_{140}^+$. Our results are in good agreement with
previous measurements.
\end{abstract}


\maketitle

\section{\label{sec:intro}Introduction}

Although melting is a universal property of bulk matter, no
quantitative theory of this phenomenon is known. A possible way to
tackle the problem is to study small systems, expecting that the
number of degrees of freedom is low enough for the theory to be
tractable. On the other hand, this makes comparison to experiments
harder, since melting of very small particles cannot be seen
directly. It has been shown however, that caloric curves could
theoretically \cite{Berry1987,Jortner1989,Labastie1990} and
experimentally \cite{Hab1998,Hab2005} be used to characterize the
melting transition. Generally, the melting temperatures of clusters
are much lower than in the bulk, but higher temperatures have been
reported for tin \cite{Shvartsburg2000} and for gallium
\cite{Jarrold2003}. The most thorough study on small clusters is
that of Schmidt \em et al \em \cite{Hab1998}. They found that the
melting temperature of small sodium particles (in the range of a
hundred atoms) is far from monotonous as a function of size. The
latent heat per atom also varies considerably. Those variations
originate from a complex interplay between electronic, geometric and
entropic effects \cite{Hab2005,Aguado2005}.

The general scheme for cluster calorimetry is the following.
Clusters are produced at a given temperature $T$ using a buffer gas
as a heat bath. They are mass selected and then some energy $E$ is
added to the clusters in order to bring them to a known reference
state. An increase $\delta T$ of the temperature is compensated by a
decrease $\delta E$ of the energy needed to reach the same state.
The heat capacity is then deduced as the ratio $C(T)=\delta E/\delta
T$. A peak in the curve C(T) is the signature of a phase transition.
The maximum of this peak gives the melting temperature and its area
is the latent heat \cite{Jortner1989}.

In the existing experimental methods, the energy $E$ is brought
either by a laser \cite{Hab1998} or by collisions
\cite{Jarrold2003}. The method of Ref. \onlinecite{Hab1998} is
accurate but requires a laser excitation. If photoabsorbtion is very
efficient for metallic clusters, this is not the case for all kind
of nanoparticles: most of them often undergo direct
photodissociation rather than heating. This drawback can be overcome
by the collisional technique \cite{Jarrold2003}. There is price to
pay however because the determination of the amount of energy
transferred by inelastic collisions relies on a model.

Another method, which is not based on the determination of the
caloric curve, has been used to extract melting temperatures
\cite{Breaux2005a}: one measures the mobility of clusters in a drift
tube, from which collision cross sections are deduced. A variation
of the cross section as a function of the temperature is the
signature of the phase transition. This method is the only one able
to detect a second-order phase transition. Nevertheless, it relies
on a subtle effect that might be hard to detect. Since we can also
measure collision cross sections in our experiment
\cite{Chirot2006}, we should also be able to detect the transition
in a similar way, but preliminary measurements have not yet revealed
any evidence of second order phase transition in sodium clusters.

This paper reports a new way to acquire caloric curves of mass
selected clusters. Basically, as in previous experiments, we take
advantage of the relation between evaporation rate and internal
energy: the reference state, hence its internal energy, corresponds
to a given evaporation rate. In our case the energy is brought by
\emph{sticking} collisions. In this way, unlike in
Ref.~\onlinecite{Jarrold2003}, the energy is increased at each
collision by a \textit{well defined} quantity $E_c+D$, where $E_c$
is the collision energy and $D$ the dissociation energy. Moreover,
as our method does not require laser excitation, it is easily
transferable to many different systems. Furthermore, our method is
parameter free: performing the experiment at two different collision
energies allows us to get rid of the unknown dissociation energies
in the final expression of the heat capacity. In this paper, we
demonstrate the validity of our method using sodium clusters, in
order to compare our results with those of
Ref.~\onlinecite{Hab2005}.

In the following, we first introduce the principle of the method
used for extracting the caloric curves (Section~\ref{sec:principe}),
then we briefly describe the experimental setup
(Section~\ref{sec:exp}). The evolution of the number of sticking in
the multicollision regime is discussed in
section~\ref{sec:evolution}. Section~\ref{sec:method} is devoted to
a detailed description of the method whose validity and robustness
is discussed with the help of Monte Carlo simulations in
section~\ref{sec:monte_carlo}. Finally, in
section~\ref{sec:exemple}, we show the experimental caloric curves
obtained for $Na_{90}^+$ and $Na_{140}^+$ and compare them to
previous results.

\section{\label{sec:principe}Principle of the method}

When performing calorimetry with
nanoclusters, there is no way to measure the temperature
\emph{after} adding energy to the system, in contrast to macroscopic systems,
where the temperature can be monitored as a function of added energy. Only the
initial temperature can be fixed by a suitable preparation of the system.
Another difficulty is related to the fact that measurements are performed on an
assembly of clusters with some statistics involved. In order to illustrate the
principle of the method, we neglect all statistical effects here. Further
details are given in section~\ref{sec:method}.

So we assume that all the clusters have initially a given internal
energy $E_0(T)$. The purpose of the experiment is to add a known
amount of energy $\Delta E$ to the cluster such as to bring it to a
well defined and recognizable state. The energy $E_f = E_0(T) +
\Delta E$ of this state is not necessarily known. All we need is
that $E_f$ does not depend on the partition between $E_0$ and
$\Delta E$. In our experiment, this final state is given by the
condition that its lifetime is equal to the detection time in the
experiment. Other experiments (see refs.
\onlinecite{Hab1998,Jarrold2003}) use slight variations of this
condition. The fact that this state is well defined is given by the
theory of the evaporative ensemble\cite{Klots1987}: since the
lifetime with respect to evaporation depends exponentially on the
energy, fixing the lifetime, even loosely, gives a well-defined
value to the energy.

\begin{figure}
\includegraphics[width=8cm]{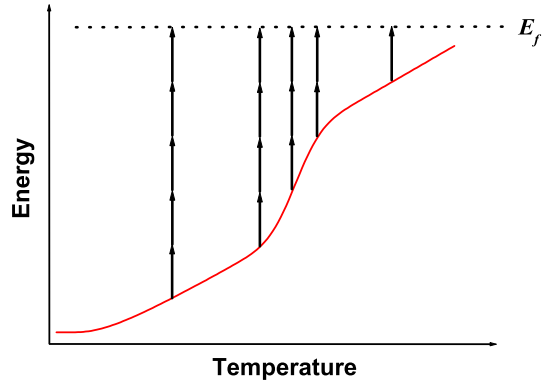}
\caption{\label{fig:courbe_capa_coll} (Color on line) Illustration
of the principle of our experiment. The cluster energy is plotted as
a function of the initial temperature. Each small vertical arrow
represents the quantity $D_n + E_c$. $E_f$ is the energy such that
the dissociation time equals the observation time in our experiment.
As the initial temperature is increased, less and less sticking are
observed. This decrease in the number of sticking evolves more
rapidly as a phase transition occurs.}
\end{figure}

There are various ways to pour energy into the cluster. Haberland's
group uses photons\cite{Hab1998}, Jarrold's uses inelastic
collisions with a rare gas\cite{Jarrold2003}. In our group, we use
sticking collisions. In this case, the size $n$ is increased by one
at each collision and the added energy is $D_{n+1}+E_c$, where $D_n$
is the dissociation energy of the cluster of size $n$ and $E_c$ the
collision energy in the center of mass frame (it is easily shown
that the rotational energy can be neglected). Basically, the energy
$E_f$ is reached when no more sticking collision is possible, that
is, the maximum number of sticking collisions depends on the initial
energy. Fig.~\ref{fig:courbe_capa_coll} illustrates this
schematically. The solid line represents the internal energy
$E_0(T)$ of the cluster as a function of its temperature. A first
order phase transition shows up  as a smooth jump in this curve
(which gives a peak in the heat capacity). The vertical arrows in
the figure represent the quantities $D_n+E_c$. $E_f$ is depicted as
constant, although it depends slightly on the size reached after the
stickings have occurred. As the initial temperature is increased,
the cluster can stand less and less sticking. Five situations are
depicted where the number of sticking evolves from 5 to 1. Before
and after the phase transition, the number of sticking evolves
slowly. But at the phase transition it evolves much more rapidly due
to the rapid change in energy as the temperature increases.

Of course, the scheme presented on Fig.~\ref{fig:courbe_capa_coll}
is oversimplified. The main correction is that even if the initial
energy is supposed to be well defined, the number of sticking
collisions results from a statistical process, with Poisson
probability. Furthermore, it is impossible to rule out evaporation
during the process. For the sake of simplicity however, let us
assume for the present discussion that we can extract from the
experiment a well defined ``maximum number of sticking'' $n_{max}$
for a given initial temperature. It is then theoretically possible
to obtain the initial internal energy $E_0$ of a cluster of initial
size $j$ as:

\begin{equation}
E_0 = E_f - \sum_{i=1}^{n_{max}}\left(D_{j+i} + E_c\right)
\end{equation}

\begin{figure}
\includegraphics[width=8cm]{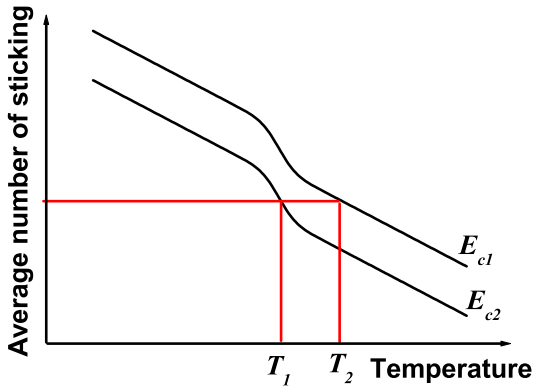}
\caption{\label{fig:illustrT1T2} (Color on line) Schematic view of
the average number of sticking as a function of the temperature. Two
curves are shown corresponding to two energies of collision
$E_{c_1}$ and $E_{c_2}$, with $E_{c_2}>E_{c_1}$. When the collision
energy is increased at a given temperature less sticking occurs.}
\end{figure}

This equation cannot be used directly however, because neither $E_f$
nor the $D_n$'s are known. We get rid of this problem by using a
differential method. The basic idea is to measure the number of
sticking $n_{max}(T,E_{c_1})$ and $n_{max}(T,E_{c_2})$ as a function
of the initial temperature $T$ for two different collision energies
$E_{c_1}$ and $E_{c_2}$, and to find two temperatures $T_1$ and
$T_2$ such that $n_{max}(T_1,E_{c_1})=n_{max}(T_2,E_{c_2})$ (see
Fig.~\ref{fig:illustrT1T2}). Then, $E_f$ and the $D_n$'s are the
same in both processes. This can be written as:

\begin{eqnarray}
E_f &=& E_0(T_1) + \sum_{i=1}^{n_{max}}\left(D_{j+i} +
E_{c_1}\right)
\label{eq:similidiff1}\\
    &=& E_0(T_2) + \sum_{i=1}^{n_{max}}\left(D_{j+i} + E_{c_2}\right)
\label{eq:similidiff2}
\end{eqnarray}

In a first approach, one can neglect the variation of the
collisional energies with the number of sticking. By subtracting
Eq.~\ref{eq:similidiff1} to Eq.~\ref{eq:similidiff2} and using a
finite differential approximation, one can derive:

\begin{equation}
\frac{\partial E_0}{\partial T} \approx n_{max}
\frac{E_{c_1}-E_{c_2}}{T_2-T_1} \label{eq:C(T)approx}
\end{equation}

Let us recall that this equation is a rough approximation however.
It is useful for the understanding of the principle of the
differential method, but it has been established with a number of
simplifying hypotheses. Specially, the quantity $n_{max}$ is not
well defined. Instead, we will show in section \ref{sec:evolution}
that a well defined and relevant quantity is the \emph{average}
number of sticking. This quantity can be easily extracted from the
experimental mass spectra after sticking.

\section{\label{sec:exp}Experimental setup}

\begin{figure}
\includegraphics[width=8cm]{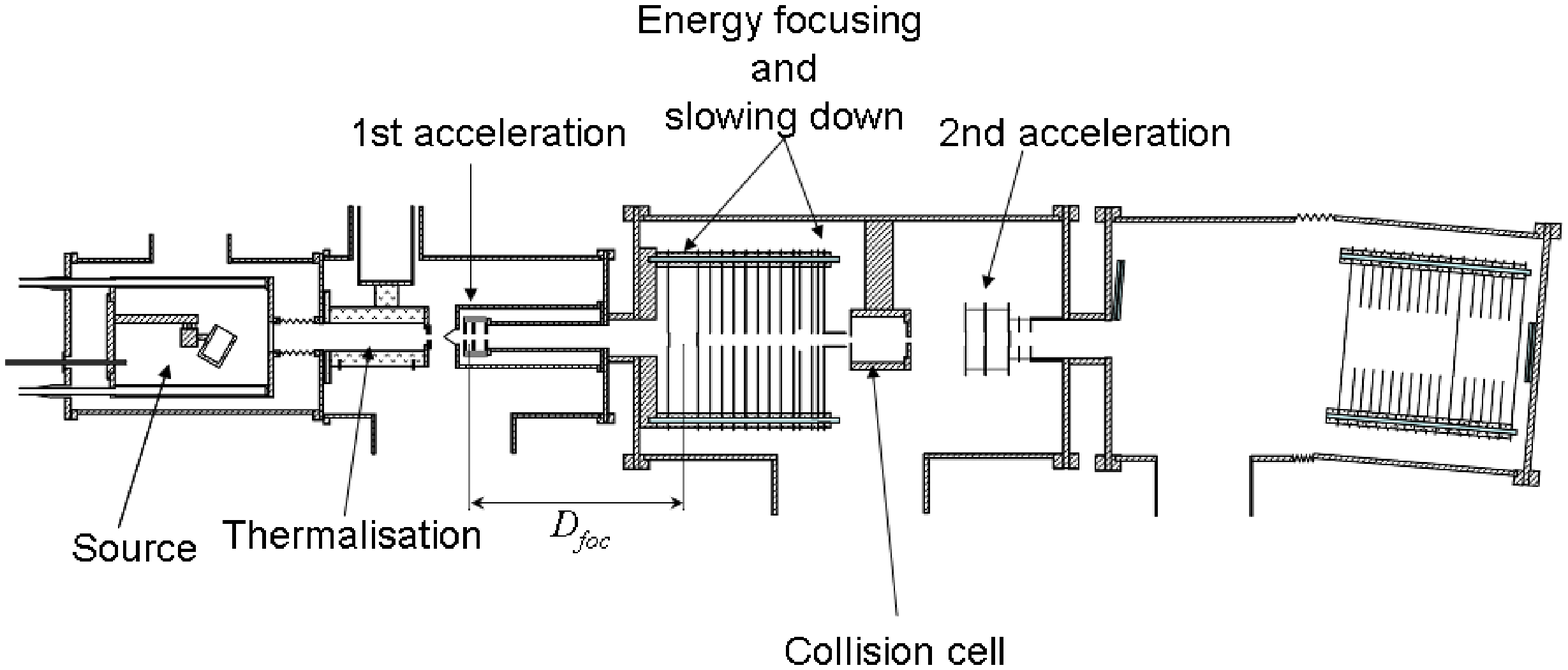}
\caption{\label{fig:exp_setup} Schematic view of the experiment.}
\end{figure}

The experimental setup is depicted in Figure~\ref{fig:exp_setup} and
has already been described in details elsewhere \cite{Chirot2006}.
The key points of the experiment are highlighted in the following.

The clusters are produced in a gas aggregation source. A crucible
filled with sodium is heated up to 573~K. The source walls are
cooled down with liquid nitrogen. A controlled flux of cold Helium
(173~K) flows through the hot sodium vapor and condensation occurs.
A hollow cathode discharge in the crucible ionizes the produced
clusters.

The helium buffer gas brings the sodium clusters to the next stage
of the experiment, the thermalization chamber. Its temperature can
be set between 140~K and 500~K. Clusters thermalization occurs
thanks to collision with the helium buffer gas ($\approx 10^5$
collisions). Out of the thermalization chamber flows a continuous
beam of charged, temperature controlled, sodium clusters. The size
distribution is controlled by varying the helium gas flow, the
temperature of the crucible and the output diameter of the
thermalization chamber. Typically, the center of the size
distribution ranges from 30 to 300 atoms with a width of about 100.

We then apply pulsed voltages to electrodes in order to mass select,
energy focus and slow down the clusters. The first acceleration
stage consisting in 3 electrodes serves two purposes: first it
allows a first raw mass selection, but more importantly, it is used
to bring the clusters at the distance $D_{foc}$ with a linear
relation between their kinetic energy and their position. Clusters
of different sizes will reach $D_{foc}$ at different times. By
applying a voltage on the energy focusing electrode at a specific
time $T_{foc}$, we compensate the energy dispersion for a given size
of clusters, giving all of them the same energy to a very good accuracy.

Mass selected clusters are then decelerated with a gradual potential
barrier until they reach the end of the slowing down apparatus. The
voltages applied on electrodes are then suddenly shut down.

At this point, clusters are mass selected, thermalized and slowed
down to a known controlled kinetic energy. They enter a cylindrical
collision cell (5~cm long, 2.5~cm radius, 5~mm entrance and exit
holes). The vapor density $\rho$ of sodium atoms in the cell is
controlled by the temperature $T_{cel}$. After colliding, clusters
finally reach the second acceleration stage, which, used in
combination with a reflectron, gives us the size distribution at the
output of the cell.

Clusters can be routinely slowed down to $E_k=10$~eV. The full width
at half maximum of the kinetic energy distribution is about 2~eV.
The mean collision energy in the center of mass frame is given by
\begin{equation}
E_c = \frac{E_k}{n+1} + \frac{3 n k_B T_{cel}}{2(n+1)}.
\label{eq:Ecoll}
\end{equation}
where the second term is the contribution from the mean kinetic
energy of the atoms in the vapor.

Our raw experimental data are the mass spectra recorded at the
output of the cell, after parent clusters have undergone collisions.
The mass spectra are recorded as a function of several parameters
such as the initial temperature of the clusters, the density in the
cell or the kinetic energy $E_k$.

The pulsed voltages are operated at 200~Hz, and mass spectra are
generally averaged over 6000 sweeps. The thermalizer or cell
temperature is ramped (1~K/minute) and mass spectra are continuously
acquired with the following sequence:
\begin{itemize}
   \item[-] Mass spectrum at $E_{c_1}$
   \item[-] Background spectrum at $E_{c_1}$
   \item[-] Mass spectrum at $E_{c_2}$
   \item[-] Background spectrum at $E_{c_2}$
 \end{itemize}
until the final temperature is reached. Background peaks originate
from masses close to the selected one, which, although not ideally
transmitted through the energy focusing device, can also cross the
collision cell and be detected. These peaks are present in the
absence of sodium vapor in the cell. Background spectra are obtained
by slightly changing $T_{foc}$ so that neither the chosen mass nor
any other mass is energy focused.

In the following section we analyze the evolution of the mass
distributions at the output of the cell as the experimental
parameters are varied.

\section{\label{sec:evolution}Mass distributions in the multicollision regime}

\begin{figure}
\includegraphics[width=8cm]{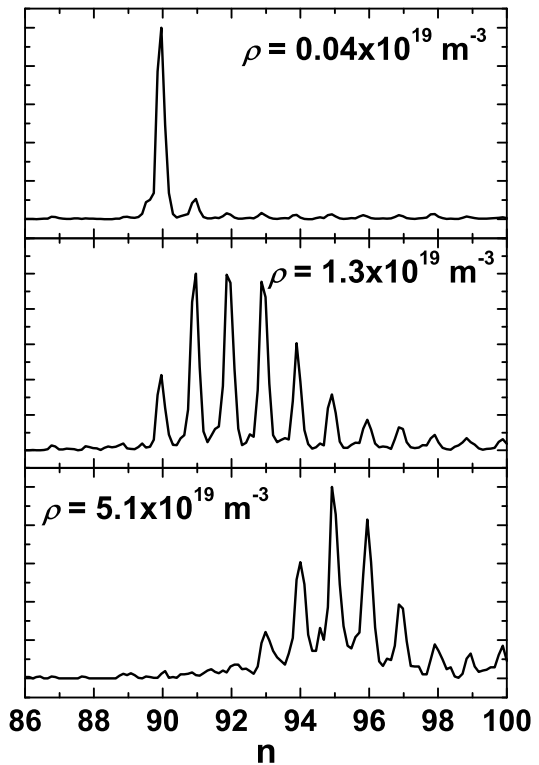}
\caption{\label{fig:spectres_collage} Size distributions for
$Na_{90}^+$ clusters going through the cell with $T=146$~K and
$E_k=20$~eV. As the density in the cell increases (from top to
bottom), the number of sticking onto $Na_{90}^+$ increases.}
\end{figure}

When the thermalized, mass selected clusters fly through the cell,
they undergo a number of collisions, depending on the density of
atoms in the cell. If the collision energy is low enough, those
collisions lead to the sticking of sodium atoms onto the clusters
\cite{Chirot2007}. Fig.~\ref{fig:spectres_collage} shows the
evolution of the size distribution at the exit of the cell when the
atom density $\rho$ is increased. These spectra are obtained for
$Na_{90}^+$ slowed down to $E_k=20$~eV and thermalized at a
temperature $T=146$~K. The top panel, for almost no density in the
cell, demonstrates the ability of our setup to mass select the
clusters $Na_{90}^+$. As can be seen, peaks appear after the
selected mass. Particular care has been taken in order to eliminate
these background peaks. Then, as the density is increased, bigger
sizes appear. Since the collision energy is low (0.25~eV in the
center of mass frame) compared to the dissociation energy (about
1~eV) and the clusters have a low temperature, several stickings are
observed (up to 8 in the lower panel of
Figure~\ref{fig:spectres_collage}).

Starting from a cluster of size $j$, a convenient way to quantify the number of
stickings, is to use the
size distribution barycenter $\overline{n}$. It is calculated as:
\begin{equation}
\label{eq:bary} \overline{n} = \frac{\sum_i i\times I(i)}{\sum_i
I(i)}
\end{equation}
where $I(i)$ is the integrated signal of clusters of size $i+j$.

\begin{figure}
\includegraphics[width=8cm]{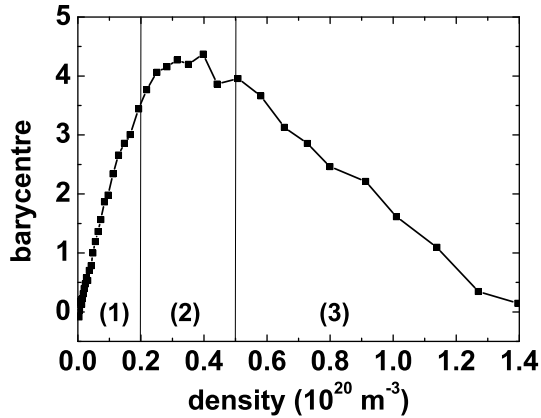}
\caption{\label{fig:nbrecollagevsTcell} Evolution of the sticking
distribution barycenter as a function of the density of sodium atoms
in the cell for $Na_{90}^+$, $T=146$~K and $E_k=20$~eV. Three
regimes can be distinguished: (1) linear increase, (2) saturation
and (3) decrease.}
\end{figure}

Fig.~\ref{fig:nbrecollagevsTcell} shows the evolution of
$\overline{n}$ as a function of the vapor density in the cell for
$Na_{90}^+$ with $E_k=20$~eV and $T=146$~K. The barycenter first
increases until it reaches a plateau at $\overline{n} \approx4.5$
for a density of 0.3$\times$10$^{20}$~m$^{-3}$. It then decreases
continuously and can even reach negative values. Two competitive
reactions occur in the cell that govern the evolution of the
size distribution, the sticking of an atom onto the cluster:
\begin{eqnarray}
Na_n^+ + Na \rightarrow Na_{n+1}^+\\
E = E_0 + D_{n+1} + E_c
\end{eqnarray}
and the evaporation:
\begin{eqnarray}
Na_n^+ \rightarrow Na_{n-1}^+ + Na\\
E = E_0 - D_n - \varepsilon
\end{eqnarray}
where $E$ and $E_0$ are the internal energies after and before
reaction respectively, $D_n$ are the dissociation energies of
cluster of size $n$, $E_c$ is the collision energy in the center of
mass frame and $\varepsilon$ the fragment kinetic energy. Three
regimes can be distinguished:

\begin{enumerate}
\item The number of sticking increases linearly with the density.
Each collision leads to a sticking and $E$ remains low enough so
that no evaporation occurs.
\item The number of sticking
reaches a maximum (saturation). At each collision, the internal
energy is increased by the dissociation energy $D_n$ plus the
collision energy $E_c$. When the internal energy gets high enough,
evaporation sets in and counterbalances the sticking: the mean
evaporation time is nearly equal to the mean time between two
collisions.
\item After saturation is reached, each extra collision leads to evaporation.
Sticking-evaporation cycles increase the internal energy of the
clusters by $E_c-\varepsilon$ ($\varepsilon\approx k_BT$). Since
$\varepsilon < E_c$, after a number of evaporation-sticking cycles,
the clusters heat up and evaporation dominates.
\end{enumerate}

The reliability of our method rests on the fact that the state of
energy $E_f$ on Fig.~\ref{fig:courbe_capa_coll} can be reached in
the course of the flight through the cell. Thus, the density has to
be high enough so that regime (2) or (3) occurs. This implies that
evaporation occurs also during the process, so that the analysis
leading to Eq.~(\ref{eq:C(T)approx}) has to be refined in order to
account for evaporation. This is done in the next section.

\section{\label{sec:method}Extraction of the caloric curves}

\subsection{Mathematical expression for the heat capacity}
Let us consider a cluster of initial energy $E_0$ and size $j$ when
it enters the cell. Suppose now that it undergoes $i$ collisions
during its flight through the cell, and that the final observed
number of stickings is $n$, with $n <i$ because some evaporation
occurs. Our basic hypothesis is that no matter the detailed history
of these events, the final value of $n$ depends only on the total
energy $E$ which has been brought to the cluster:
\begin{equation}
   E=E_0+iE_c+f
\end{equation}
where $f$ is an unknown quantity, which depends on the various
values of the dissociation energies but neither on $E_0$ nor on
$E_c$. There is no need in knowing $f$, as it does not appear in the
final expression of the heat capacity. This hypothesis amounts to
say that the number of evaporations is such as to bring back the
cluster to the energy $E_f$ defined in section~\ref{sec:principe},
and thus depends only on the excess energy $E-E_f$.

Now, all three quantities $E_0$, $E_c$ and $i$ are statistically distributed.
In this treatment, we suppose that the widths of the $E_0$ and $E_c$
distributions are small respective to their averages, so that $E_0$ can be
replaced by its average $E_0(T)$ and $E_c$ by its average Eq. (\ref{eq:Ecoll}).
The main statistical effects come from the distribution of the number of
collisions $i$, whose width is close to the average. The probability $P(i)$ is
a Poisson law~:
\begin{equation}
   P(i)=\frac{\overline{\imath}^ie^{-\overline{\imath}}}{i!}
\label{eq:Poisson}
\end{equation}
with an average $\overline{\imath}$, which can be easily computed knowing the
density $\rho$ in the cell.

Computing the mean number of sticking, we obtain:
\begin{equation}
  \overline{n}=\sum_i P(i)n(E_0(T)+iE_c).
\end{equation}
By the implicit functions theorem, we have:
\begin{equation}
   \left.\frac{\partial E_0}{\partial E_c}\right|_{\overline{n}}=
-\frac{\partial\overline{n}}{\partial E_c}
\bigg/\frac{\partial\overline{n}}{\partial E_0} \label{eq:implicit}
\end{equation}
The two derivatives on the r.h.s. above are given by:
\begin{eqnarray}
  \frac{\partial\overline{n}}{\partial E_0}&=&
\sum_i P(i)\frac{\partial n}{\partial E}(E_0(T)+iE_c)\label{eq:dE0}\\
  \frac{\partial\overline{n}}{\partial E_c}&=&
\sum_i iP(i)\frac{\partial n}{\partial E}(E_0(T)+iE_c)\label{eq:dEc}
\end{eqnarray}
We now use a specific property of the Poisson law:
\begin{equation}
   iP(i)=\overline{\imath}\Big(P(i)+\frac{\partial
P(i)}{\partial\overline{\imath}}\Big)
\end{equation}
so that Eq.~(\ref{eq:dEc}) can be rewritten as:
\begin{equation}
  \frac{\partial\overline{n}}{\partial E_c}=
\overline{\imath}\left(\frac{\partial\overline{n}}{\partial
E_0}+\frac{\partial^2\overline{n}}{\partial
E_0\partial\overline{\imath}}\right)\label{eq:dEc_finale}
\end{equation}

So far, we neglected the change in cluster kinetic energy after
each collision. When the number of collisions becomes large, this
has to be accounted for through an effective value $E^*_c$, which is
to a good approximation:
\begin{equation}
E_{c}^{*} = E_c\frac{j+1}{\overline{\imath}}
\left(1-\left(\frac{j}{j+1}\right)^{\overline{\imath}}\right)
\end{equation}
We use now
\begin{equation}
C(T)=\frac{dE_0}{dT}
\end{equation}
together with equations (\ref{eq:implicit} and
(\ref{eq:dEc_finale}), and we eventually obtain the final expression
for the heat capacity:
\begin{equation}
\label{eq:C(T)_finale} C(T) = -
\overline{\imath}\left(1+\frac{\partial^2\overline{n}}{\partial
T\partial\overline{\imath}}\bigg/\frac{\partial
\overline{n}}{\partial T}\right)\left.\frac{\partial E_c^*}{\partial T}
\right|_{\overline{n}}
\end{equation}

We describe in the next section how the different terms in
Equation~(\ref{eq:C(T)_finale}) are deduced from the experiment.

\subsection{Practical extraction of $C(T)$ from experimental curves}

In order to use Eq.~(\ref{eq:C(T)_finale}), we need three partial derivatives
and the value of $\overline{\imath}$. All these quantities can be extracted
from experiment.
\subsubsection{The derivative $\displaystyle\left.\frac{\partial E_c}{\partial
T}\right|_{\overline{n}}$} The method can be summarized as follows
(see Fig.~\ref{fig:principeT1T2} for an illustration):
\begin{figure}
\includegraphics[width=8cm]{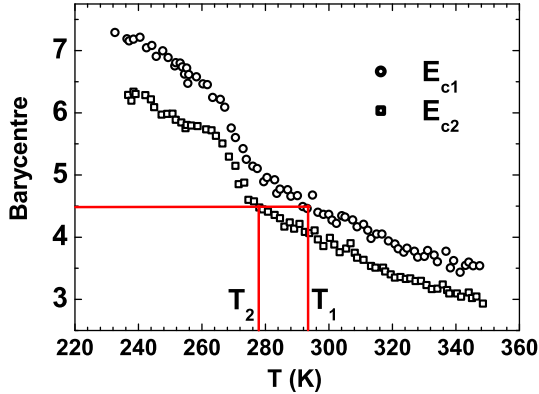}
\caption{\label{fig:principeT1T2} (Color on line) Experimental data
showing the barycenter $\overline{n}$ as a function of the initial
temperature for $Na_{140}^+$ clusters. The two curves correspond to
collision energies of $E_{c_1}=0.14$~eV and $E_{c_2}=0.21$~eV}
\end{figure}

\begin{enumerate}
\item We record the barycenter $\overline{n}$ of the size distribution as a
function of the initial cluster temperature for two different
kinetic energies of the clusters $E_{c_1}$ and $E_{c_2}$.

\item We then chose a given value of the barycenter. For this value,
we record the two corresponding temperatures $T_1$ and $T_2$ (as shown on
Fig.~\ref{fig:principeT1T2}).

\item We approximate:
\begin{equation}
   \left.\frac{\partial E_c}{\partial T}\right|_{\overline{n}}
=\frac{E_{c_2}-E_{c_1}}{T_2-T_1}
\end{equation}

\end{enumerate}

\subsubsection{Determination of $\overline{\imath}$}
The average number of collisions $\overline{\imath}$ can be
considered to be the same for the two collision energies $E_{c_1}$
and $E_{c_2}$, as long as the difference $\delta E_c$ remains small
enough (in other words, that the method is indeed differential). For
instance, for clusters as small as $Na_{30}^+$ at $E_k=15$~eV and
20~eV (that correspond to 0.53 and 0.70~eV collision energies) and
for a density $\rho=0.2\times10^{20}$~m$^{-3}$ in the cell, Monte
Carlo simulations (see next section) show that the average number of
collisions are respectively 2.11 and 2.08. As one can see, the
difference is small enough that it can be neglected. Note that as
the cluster size increases it becomes even easier to fulfill this
requirement.

\begin{figure}
\includegraphics[width=8cm]{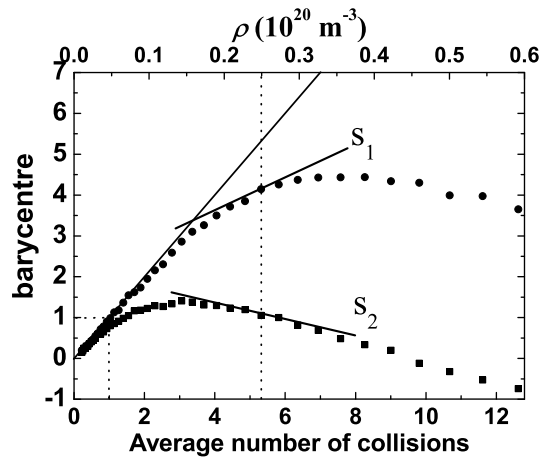}
\caption{\label{fig:determinationh} Monte Carlo simulation of
$\overline{n}(\rho)$ for two initial temperatures of $Na_{90}^+$
clusters: 146~K (circles) and 293~K (squares) (see
Section~\ref{sec:monte_carlo} for details). The kinetic energy of
the cluster is $20$~eV. The bottom scale is in average number of
collisions $\overline{\imath}$ whereas the top scale is in vapor
density. At low density, the number of sticking is equal to the
number of collisions. $S_1$ and $S_2$ are the slopes used in the
determination of
$\frac{\partial^2\overline{n}}{\partial\overline{\imath}\partial
T}$.}
\end{figure}

The average number of collisions $\overline{\imath}$ is deduced from
saturation curves such as the ones presented in
Figure~\ref{fig:determinationh}. There is no need in knowing the
value of the cross sections to determine this parameter. The main reason is
that in regime (1), each collision leads to a sticking. The proportionality
factor between the cell density $\rho$ and $\overline{\imath}$ is thus deduced
from the initial slope of the curves. Actually collisions at high impact
parameter might not lead to sticking. These collisions involve a very small
energy exchange anyway, so that they can be neglected\cite{Chirot2007}.

\subsubsection{The derivatives
$\displaystyle\frac{\partial^2\overline{n}}{\partial
T\partial\overline{\imath}}$ and $\displaystyle\frac{\partial
\overline{n}}{\partial T}$}
The principle is presented on Fig.~\ref{fig:determinationh} too. The slopes
$S_1$ and $S_2$ are the derivatives
$\frac{\partial \overline{n}}{\partial \overline{\imath}}$ at
temperatures $T_1$ and $T_2$ respectively. $\frac{\partial^2
\overline{n}}{\partial T
\partial \overline{\imath}}$ is then estimated as $(S_2-S_1)/(T_2-T_1)$. Note
that its value
does not depend on the collision energy (which needs nevertheless to
be the same for the two curves at $T_1$ and $T_2$).

Finally the derivative $\frac{\partial \overline{n}}{\partial T}$ is
evaluated using the curves $\overline{n}(T)$ (see
Fig.~\ref{fig:principeT1T2}).

\section{\label{sec:monte_carlo}Monte Carlo simulation: Validation of the method}

\begin{figure}
\includegraphics[width=8cm]{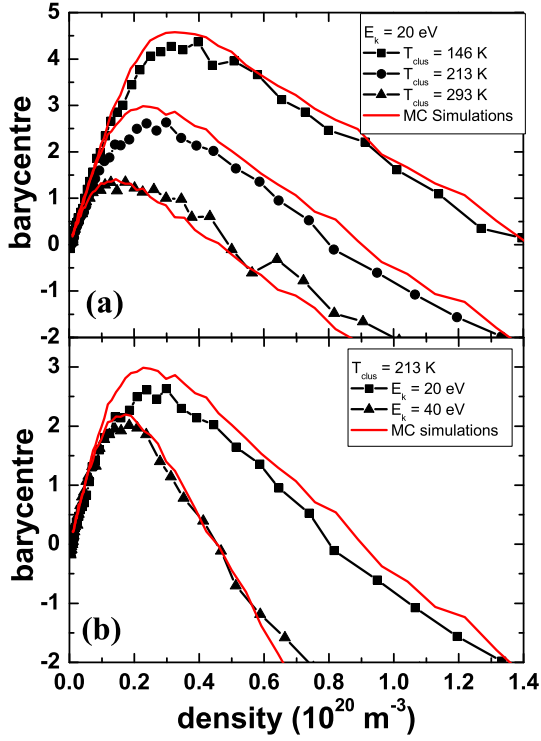}
\caption{\label{fig:n=90_ncolvsTcell_Tini} (Color on line) Evolution
of the size distribution barycenter as a function of the density of
sodium atoms in the cell for $Na_{90}^+$ for (a) three different
initial temperatures (the kinetic energy is $E_k=20$~eV) and (b) two
different kinetic energies ($T=213$~K). Full lines are the results
of Monte Carlo simulations (see Section~\ref{sec:monte_carlo}).}
\end{figure}

\begin{figure}
\includegraphics[width=8cm]{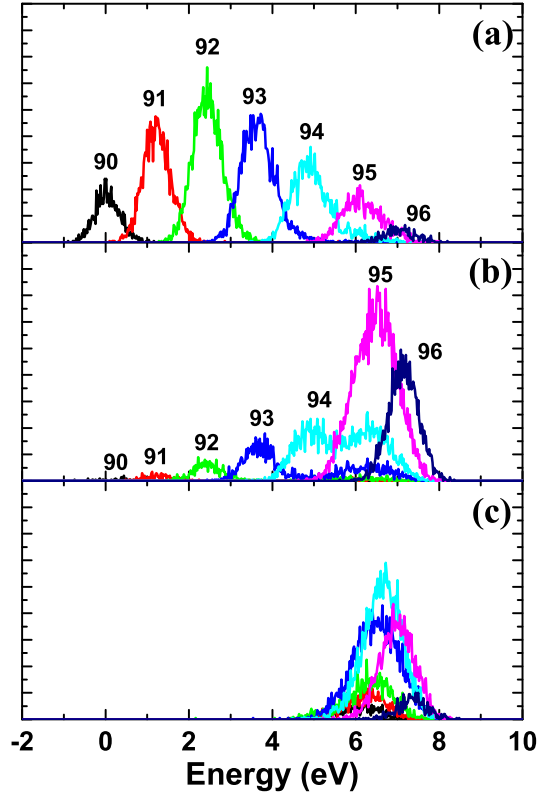}
\caption{\label{fig:n=90_distrinrj_interne} (Color online)
Simulation of the internal energy distribution for three different
densities in the cell: (a) 0.1, (b) 0.32 and (c) 0.72$\times10^{20}$
m$^{-3}$. The initial temperature of the clusters is 146~K and
$E_k$=20~eV. Peaks are labeled by the size of the clusters.}
\end{figure}
\subsection{Description of the simulations}
Realistic Monte Carlo simulations have been performed in order to
check the validity of the method under real conditions where the
above analytical ideal approach might be invalidated by a number of
details that are not taken into account: initial internal and
collisional energy distributions, spatial divergence due to
collisions, variations of $\frac{\partial^2 \overline{n}}{\partial T
\partial \overline{\imath}}$, dissociation events between the output of the cell
and
the detector,...

The initial internal energy of the clusters is randomly picked such
that the average energy is equal to:
\begin{equation}
\langle E\rangle=\frac{(3j-6)h\nu_0}{e^{\frac{h\nu_0}{k_BT}}-1}
\end{equation}
where $j$ is the number of atoms in the cluster, $h$ the Planck
constant, $\nu_0$ the vibration frequency. The frequency $\nu_0$ is
taken from \onlinecite{nuo_sodium} and is equal to
2.3$\times$10$^{12}$~Hz. The phase transition is taken into account
by adding the latent heat $L_j(T)$ to $\langle E\rangle$ in the
following way assuming a gaussian peak in the heat capacity:
\begin{equation}
L_j(T) = L_{j0}\frac{(1-\hbox{erf}(-\frac{(T-T_{jf})}{\Delta T_j}))}{2}
\label{eq:L(T)_simul}
\end{equation}
where $T_{jf}$ is the temperature of the phase transition, $\Delta
T_j$ is the width of the phase transition and erf is the error
function. $L_{j0}$ is the latent heat for a cluster of size $j$.
This way we simulate the broadening of the phase transition for a
finite system.

The evaporation rate is estimated using a Weisskopf
approximation:
\begin{equation}
   \Gamma_{\rm evap}=8\pi\nu_0(3j-7)\sigma\frac{(1-D_j/E)^{3j-8}}{E}
\label{eq:weisskopf}
\end{equation}
where $\sigma$ is the sticking cross section to the cluster of size $j-1$.
The
collision rate is given by:
\begin{equation}
   \Gamma_{\rm coll}=\rho\sigma v_{\rm rel}
\label{eq:coll_rate}.
\end{equation}
where $v_{\rm rel}$ is the relative velocity of the cluster and the sticking
atom. The velocity
distribution of atoms is assumed to follow a Boltzman distribution.

At each collision, the change in speed of the clusters is taken
into account. Since only clusters with the right velocity are
detected, clusters that undergo a lot of collisions become too slow
to be detected and are discarded in the simulation.

We first checked that our Monte Carlo simulations could reproduce the
experimental
results of Figs.~\ref{fig:nbrecollagevsTcell} and
\ref{fig:n=90_ncolvsTcell_Tini}. The agreement between the
simulations and the experimental data is good. In order to reproduce
correctly the experimental results it is important to put in the
correct latent heat and melting temperature for $Na_{90}^+$. We took
$T_f=205$~K and $L=8$~meV/atom for the melting temperature and the
latent heat respectively. On the other hand, for the other sizes the
latent heat and temperature of fusion can be chosen arbitrarily:
there is no change in the curve $\overline{n}(\rho)$.

For the dissociation energies, only values of $D_n$ up to n=37
\cite{Brechignac1989} are available. In the simulations presented
here, the dissociation energies are all taken equal to 0.94~eV.

These simulations allow us to follow the evolution of the internal
energy distribution of the clusters as a function of their size as
the density in the cell is varied. In
Figure~\ref{fig:n=90_distrinrj_interne} we present the energy
distribution of the clusters for three different densities in the
cell. In Figure~\ref{fig:n=90_distrinrj_interne}(a) the density in
the cell is $0.1\times10^{20}$ m$^{-3}$, corresponding to an average
number of sticking $\overline{n}\approx 3$ (in this case evaporation
is completely negligible so that each collision leads to a
sticking). We see in the figures well separated peaks corresponding
to the different masses obtained after the cell. The energy
separation between the peaks correspond to the dissociation energy
plus the collision energy. The width of the peaks originates mainly
from the initial canonical energy distribution. This can be seen
from the $n=90$ energy distribution: indeed, in this example part of
the parent clusters did not collide nor evaporate. Nevertheless, as
the cluster size increases we observe an increase in the width. This
comes from the width of the collision energy distribution which adds
at each new collision.

In Figure~\ref{fig:n=90_distrinrj_interne}(b), the density is
$0.32\times10^{20}$ m$^{-3}$ ($\overline{\imath}\approx 7$). This
corresponds to the saturated case (regime (2) of
Figure~\ref{fig:nbrecollagevsTcell}). If we consider the sizes
$n=93$ and 94, the internal energy distribution becomes bimodal: the
high energy part comes from the evaporation of bigger clusters. For
the two biggest sizes, namely $n=95$ and 96, one can see as well
that the energy separation between the peaks gets smaller. In fact,
there is a strong contribution coming from the evaporation of $n=96$
clusters in the internal energy distribution of the $n=95$ clusters.
For the size $n=96$, there is of course also contribution from the
evaporation of bigger sizes that are not observed. Indeed due to
their high evaporation rate, they evaporate before detection.

As the density is increased to $0.72\times10^{20}$ m$^{-3}$
(Figure~\ref{fig:n=90_distrinrj_interne}(c)), all clusters end up
with almost the same internal energy. In this last case the average
number of collisions is about 17, much more than the maximum number
of sticking observed (which is about 6). The observed internal
energy is the maximum energy clusters can bear without evaporating
before detection. In this case the observed clusters have all
undergone evaporation.

Despite the complicated evolution of the internal energy of the
clusters as the number of collision increases, the size
distribution, for a given collision energy and density in the cell,
is still essentially governed by the internal energy of the incoming
cluster. Furthermore, as shown below, the fact that clusters of a
specific size don't have necessarily a well defined internal energy
do not prevent the determination of the caloric curves.

\begin{figure}
\includegraphics[width=8cm]{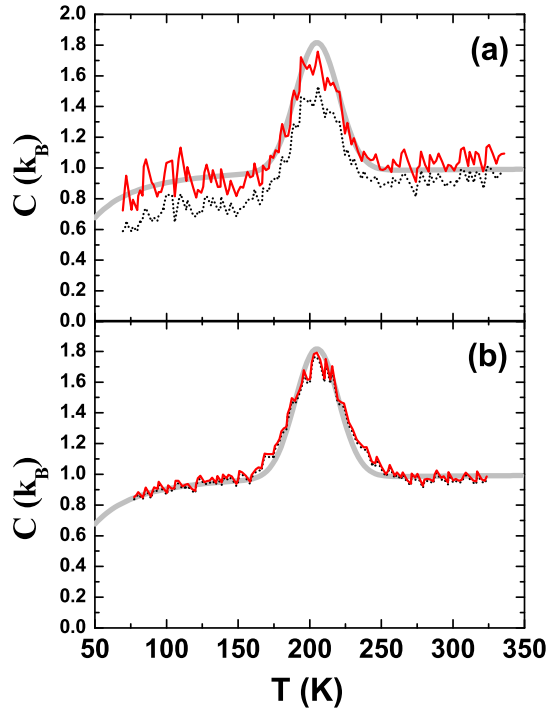}
\caption{\label{fig:courbecapafromsimul} (Color online) The thick
solid line is the caloric curve introduced in the simulation ($L$ =
8 meV/atom, $T_f$=205~K and $\Delta T$= 30~K). The dotted curves
correspond to the caloric curves obtained by setting the second
derivative in Eq.~\ref{eq:C(T)_finale} to $0$ whereas for the curves
in thin solid line it is taken into account. The top panel
correspond to $\overline{\imath}=5.3$ and the bottom one to
$\overline{\imath}=10.6$}
\end{figure}

\subsection{Test of the method of section~\ref{sec:method}}
In order to test the method of section~\ref{sec:method}, we start from a
theoretical caloric curve for $Na_{90}^+$ and we perform a set of simulations
that mimics the experimental curves. Then, using these fake experimental
curves, we apply the method and we extract a caloric curve, which is then
compared to the theoretical one we started with.

First of all, two saturation curves at $E_k=20$~eV were simulated
for two initial temperatures of the clusters, namely 146~K and
293~K. These curves are presented in Fig~\ref{fig:determinationh}
and, as already shown, allow to calculate the various terms involved in
Eq.~(\ref{eq:C(T)_finale}).

Mass spectra were then simulated at $E_c \approx 0.4$ and $0.5$~eV
as a function of the initial temperature of $Na_{90}^+$ clusters.
5000 trajectories were simulated at each temperature. Two set of
simulations were performed at two different vapor densities,
corresponding respectively to $\overline{\imath}=5.3$ and
$\overline{\imath}=10.6$. In the first case the working density
roughly corresponds to the maximum of the saturation curve (see
Fig~\ref{fig:determinationh}). It would not be a very good choice in
a real experiment since the curvature of $\frac{\partial n}{\partial
\overline{\imath}}$, thus the value of the second derivative in
Eq.~(\ref{eq:C(T)_finale}), is maximum. For
$\overline{\imath}=10.6$, the second derivative in
Eq.~(\ref{eq:C(T)_finale}) is much smaller because the saturation
curves already become much more parallel (see
Fig~\ref{fig:determinationh}).

Figure~\ref{fig:courbecapafromsimul} shows a comparison between the
theoretical caloric curve introduced in the calculation and the curves deduced
from the simulation of the experiment.
Fig.~\ref{fig:courbecapafromsimul}(a) corresponds to
$\overline{\imath}=5.3$ and Fig.~\ref{fig:courbecapafromsimul}(b) to
$\overline{\imath}=10.6$.

In order to evaluate the effect of the second derivative in
Eq.~(\ref{eq:C(T)_finale}), two curves are represented for each
vapor density in each panel of Fig.~\ref{fig:courbecapafromsimul},
respectively with and without this term. For
$\overline{\imath}=10.6$, within the noise, the curves are virtually
indistinguishable. On the other hand, for $\overline{\imath}=5.3$,
although the general shape of the caloric curve is correctly
obtained with or without the second derivative, a quantitative
agreement is obtained only by having it taken into account. The
correction nicely improves the curve even in this unfavorable case.
Nevertheless the smallest the correction the more accurate the
result. Note that the noise is more important for
$\overline{\imath}=5.3$ than for $\overline{\imath}=10.6$: since the
mean energy difference is roughly $\overline{\imath}\delta E_c$, the
two curves $\overline{n}(T)$ get closer for small
$\overline{\imath}$ values.

As already mentioned the simulations presented here are done with
all clusters having the same dissociation energies. Nevertheless, we
carefully checked that the introduction of magic numbers
(\textit{i.e.} higher dissociation energies) for arbitrary sizes
does not influence our determination of the caloric curve.

Furthermore we performed simulations with randomly chosen
temperature of fusion, latent heat and width of the transition for
cluster sizes other than the initial one, and again this does not
affect our determination of the latent heat and temperature of
fusion of the initial cluster size.

From these simulations, we find our method particularly robust. The
simulations demonstrate that the average number of sticking
$\overline{n}$ is a well suited quantity to characterize the size
distribution, even when evaporation dominates, provided
Eq.~(\ref{eq:C(T)_finale}) is used.

\section{\label{sec:exemple}Examples of experimental results with $Na_{90}^+$
and $Na_{140}^+$}

\begin{figure}
\includegraphics[width=8cm]{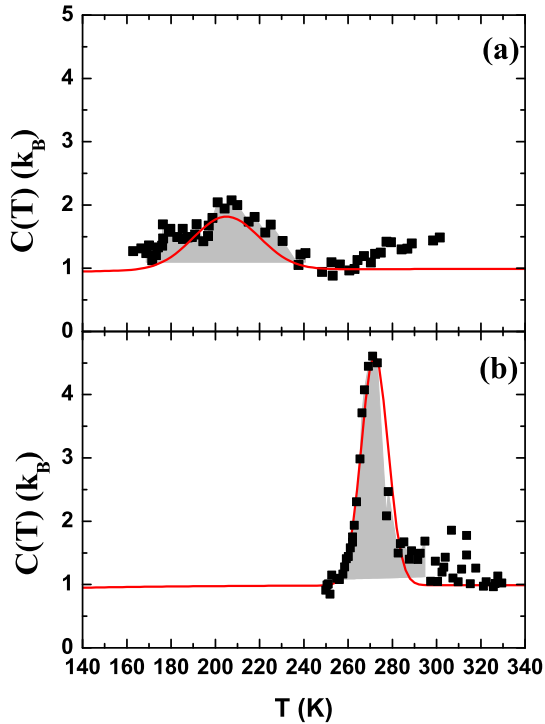}
\caption{\label{fig:Na90_140_capaexp} (Color online) Experimental
caloric curves for (a) $Na_{90}^+$ and (b) $Na_{140}^+$ clusters.
$C(T)$ is in unit of $k_B$ per degree of freedom. The full lines
represent the latent heat used in the simulations (see
Eq.~\ref{eq:L(T)_simul}) and are used to guide the eye. The grey
filled area under the curves represent the latent heat.}
\end{figure}

We present in Figure~\ref{fig:Na90_140_capaexp} the experimental
caloric curves obtained for $Na_{90}^+$ and $Na_{140}^+$ clusters.
The melting temperature is obtained from these curves as the maximum
of the peak. The latent heat is deduced by integrating the peak
area, after removing the base line (grey filled area under the
curves in figure~\ref{fig:Na90_140_capaexp}). The caloric curve for
$Na_{90}^+$ was not measured previously. We find a melting
temperature of 205~$\pm$5~K and a latent heat per atom of
8~$\pm$2~meV.

For the $Na_{140}^+$ clusters, the melting temperature is
272~$\pm$4~K and the latent heat is 14~$\pm$2~meV/atom. This is
consistent with
previous measurements by Schmidt \textit{et al}~\cite{Hab2005} ($T_{fus}=262$~K, $L=11$~meV/atom).\\
The results for $Na_{90}^+$ are quite noisy: this is due to the
relatively small latent heat of this cluster and the fact that the
width of the transition is relatively large. Nevertheless we can
still identify a peak in the caloric curve for this disfavoring
case. We observe that the latent heat increases at high
temperatures. Such an increase has already been observed, although
less pronounced (see [\onlinecite{Hab1998}] for $Na_{192}^+$ and
[\onlinecite{Kusche1999}] for $Na_{139}^+$). It might be due to
either a rising background due to a high Debye temperature or a
melting occurring in several steps. However the noise level is such
that we can not confirm the significance of this increase in our
case.

As shown from Fig.~\ref{fig:courbecapafromsimul} the exact choice of
the vapor density in the cell is not critical. Nevertheless, an
optimal choice exists. As already mentioned in
Sec.~\ref{sec:monte_carlo}, it can be chosen so that it minimizes
the noise. In order to minimize the second derivative in
Eq.~(\ref{eq:C(T)_finale}), the vapor density $\rho$ has to be
chosen high enough such that the curves $\overline{n}(\rho)$ become
almost parallel. On the other hand, if $\rho$ becomes too large,
evaporation becomes too strong and the signal gets scattered over
many mass peaks, which is unfavorable for the statistics.
Furthermore the background pressure in the collision chamber gets
too high and leads to unwanted collisions out of the cell. Finally,
too large $\overline{\imath}$ result in angular scattering of the
clusters.

The use of a differential method has several advantages. First, we
do not need to know the dissociation energies. Second, the presence
of magic numbers is not a problem with this method. Indeed, any
accident in the curves $\overline{n}(T)$ is canceled out by the
differentiation. These all in one make this method reliable and
precise. On the other hand, noise is a drawback. Since we rely on
the calculation of finite differences for the determination of the
caloric curves, and that the experimental data is discrete, the
numerical differentiation is inherently noisy.

\section{\label{sec:conclu}Conclusion}

We have developed a novel experimental method for the measurement of
caloric curves of small nanoparticles based on the sticking of atoms
onto clusters. We have mathematically established the method and
shown with the help of simulations that it is extremely robust. It
associates the accuracy of the photoexcitation methods and the large
application field of collisional methods. Moreover, its principle
makes it rigorously independent on preliminary assumptions about
properties such as dissociation energies or collisional energy
transfer.

We have experimentally demonstrated the use of the method on sodium
clusters (namely $Na_{90}^+$ and $Na_{140}^+$).

\begin{acknowledgments}

This work has been partly funded by the Agence Nationale de la
Recherche under grant ANR-05-BLAN-0145. We gratefully acknowledge M.
Schmidt for fruitful discussions.

\end{acknowledgments}

\bibliography{613841JCP}

\begin{thebibliography}{15}
\expandafter\ifx\csname natexlab\endcsname\relax\def\natexlab#1{#1}\fi
\expandafter\ifx\csname bibnamefont\endcsname\relax
  \def\bibnamefont#1{#1}\fi
\expandafter\ifx\csname bibfnamefont\endcsname\relax
  \def\bibfnamefont#1{#1}\fi
\expandafter\ifx\csname citenamefont\endcsname\relax
  \def\citenamefont#1{#1}\fi
\expandafter\ifx\csname url\endcsname\relax
  \def\url#1{\texttt{#1}}\fi
\expandafter\ifx\csname urlprefix\endcsname\relax\def\urlprefix{URL }\fi
\providecommand{\bibinfo}[2]{#2}
\providecommand{\eprint}[2][]{\url{#2}}

\bibitem[{\citenamefont{Davis et~al.}(1987)\citenamefont{Davis, Jellinek, and
  Berry}}]{Berry1987}
\bibinfo{author}{\bibfnamefont{H.-L.} \bibnamefont{Davis}},
  \bibinfo{author}{\bibfnamefont{J.}~\bibnamefont{Jellinek}}, \bibnamefont{and}
  \bibinfo{author}{\bibfnamefont{R.-S.} \bibnamefont{Berry}},
  \bibinfo{journal}{J.\ Chem.\ Phys.} \textbf{\bibinfo{volume}{86}},
  \bibinfo{pages}{6456} (\bibinfo{year}{1987}).

\bibitem[{\citenamefont{Bixon and Jortner}(1989)}]{Jortner1989}
\bibinfo{author}{\bibfnamefont{M.}~\bibnamefont{Bixon}} \bibnamefont{and}
  \bibinfo{author}{\bibfnamefont{J.}~\bibnamefont{Jortner}},
  \bibinfo{journal}{J. Chem. Phys.} \textbf{\bibinfo{volume}{91}},
  \bibinfo{pages}{1631} (\bibinfo{year}{1989}).

\bibitem[{\citenamefont{Labastie and Whetten}(1990)}]{Labastie1990}
\bibinfo{author}{\bibfnamefont{P.}~\bibnamefont{Labastie}} \bibnamefont{and}
  \bibinfo{author}{\bibfnamefont{R.~L.} \bibnamefont{Whetten}},
  \bibinfo{journal}{Phys. Rev. Lett.} \textbf{\bibinfo{volume}{65}},
  \bibinfo{pages}{1567} (\bibinfo{year}{1990}).

\bibitem[{\citenamefont{Schmidt et~al.}(1998)\citenamefont{Schmidt, Kusche, von
  Issendorf, and Haberland}}]{Hab1998}
\bibinfo{author}{\bibfnamefont{M.}~\bibnamefont{Schmidt}},
  \bibinfo{author}{\bibfnamefont{R.}~\bibnamefont{Kusche}},
  \bibinfo{author}{\bibfnamefont{B.}~\bibnamefont{von Issendorf}},
  \bibnamefont{and}
  \bibinfo{author}{\bibfnamefont{H.}~\bibnamefont{Haberland}},
  \bibinfo{journal}{Nature (London)} \textbf{\bibinfo{volume}{393}},
  \bibinfo{pages}{238} (\bibinfo{year}{1998}).

\bibitem[{\citenamefont{Haberland et~al.}(2005)\citenamefont{Haberland,
  Hippler, Donges, Kostko, Schmidt, and von Issendorf}}]{Hab2005}
\bibinfo{author}{\bibfnamefont{H.}~\bibnamefont{Haberland}},
  \bibinfo{author}{\bibfnamefont{T.}~\bibnamefont{Hippler}},
  \bibinfo{author}{\bibfnamefont{J.}~\bibnamefont{Donges}},
  \bibinfo{author}{\bibfnamefont{O.}~\bibnamefont{Kostko}},
  \bibinfo{author}{\bibfnamefont{M.}~\bibnamefont{Schmidt}}, \bibnamefont{and}
  \bibinfo{author}{\bibfnamefont{B.}~\bibnamefont{von Issendorf}},
  \bibinfo{journal}{Phys. Rev. Lett.} \textbf{\bibinfo{volume}{94}},
  \bibinfo{pages}{035701} (\bibinfo{year}{2005}).

\bibitem[{\citenamefont{Shvartsburg and Jarrold}(2000)}]{Shvartsburg2000}
\bibinfo{author}{\bibfnamefont{A.~A.} \bibnamefont{Shvartsburg}}
  \bibnamefont{and} \bibinfo{author}{\bibfnamefont{M.~F.}
  \bibnamefont{Jarrold}}, \bibinfo{journal}{Phys. Rev. Lett.}
  \textbf{\bibinfo{volume}{85}}, \bibinfo{pages}{2530} (\bibinfo{year}{2000}).

\bibitem[{\citenamefont{Breaux et~al.}(2003)\citenamefont{Breaux, Benirschke,
  Sugai, Kinnear, and Jarrold}}]{Jarrold2003}
\bibinfo{author}{\bibfnamefont{G.~A.} \bibnamefont{Breaux}},
  \bibinfo{author}{\bibfnamefont{R.~C.} \bibnamefont{Benirschke}},
  \bibinfo{author}{\bibfnamefont{T.}~\bibnamefont{Sugai}},
  \bibinfo{author}{\bibfnamefont{B.~S.} \bibnamefont{Kinnear}},
  \bibnamefont{and} \bibinfo{author}{\bibfnamefont{M.~F.}
  \bibnamefont{Jarrold}}, \bibinfo{journal}{Phys. Rev. Lett.}
  \textbf{\bibinfo{volume}{91}}, \bibinfo{pages}{215508}
  (\bibinfo{year}{2003}).

\bibitem[{\citenamefont{Aguado and Lopez}(2005)}]{Aguado2005}
\bibinfo{author}{\bibfnamefont{A.}~\bibnamefont{Aguado}} \bibnamefont{and}
  \bibinfo{author}{\bibfnamefont{J.~M.} \bibnamefont{Lopez}},
  \bibinfo{journal}{Phys. Rev. Lett.} \textbf{\bibinfo{volume}{94}},
  \bibinfo{pages}{233401} (\bibinfo{year}{2005}).

\bibitem[{\citenamefont{Breaux et~al.}(2005)\citenamefont{Breaux, Cao, and
  Jarrold}}]{Breaux2005a}
\bibinfo{author}{\bibfnamefont{G.}~\bibnamefont{Breaux}},
  \bibinfo{author}{\bibfnamefont{B.}~\bibnamefont{Cao}}, \bibnamefont{and}
  \bibinfo{author}{\bibfnamefont{M.}~\bibnamefont{Jarrold}},
  \bibinfo{journal}{J. Phys. Chem. B} \textbf{\bibinfo{volume}{109}},
  \bibinfo{pages}{16575} (\bibinfo{year}{2005}).

\bibitem[{\citenamefont{Chirot et~al.}(2006)\citenamefont{Chirot, Zamith,
  Labastie, and L'Hermite}}]{Chirot2006}
\bibinfo{author}{\bibfnamefont{F.}~\bibnamefont{Chirot}},
  \bibinfo{author}{\bibfnamefont{S.}~\bibnamefont{Zamith}},
  \bibinfo{author}{\bibfnamefont{P.}~\bibnamefont{Labastie}}, \bibnamefont{and}
  \bibinfo{author}{\bibfnamefont{J.-M.} \bibnamefont{L'Hermite}},
  \bibinfo{journal}{Rev. Sci. Instrum.} \textbf{\bibinfo{volume}{77}},
  \bibinfo{pages}{063108} (\bibinfo{year}{2006}).

\bibitem[{\citenamefont{Klots}(1987)}]{Klots1987}
\bibinfo{author}{\bibfnamefont{C.}~\bibnamefont{Klots}}, \bibinfo{journal}{Z.\
  Phys.\ D} \textbf{\bibinfo{volume}{5}}, \bibinfo{pages}{83}
  (\bibinfo{year}{1987}).

\bibitem[{\citenamefont{Chirot et~al.}(2007)\citenamefont{Chirot, Zamith,
  Labastie, and L'Hermite}}]{Chirot2007}
\bibinfo{author}{\bibfnamefont{F.}~\bibnamefont{Chirot}},
  \bibinfo{author}{\bibfnamefont{S.}~\bibnamefont{Zamith}},
  \bibinfo{author}{\bibfnamefont{P.}~\bibnamefont{Labastie}}, \bibnamefont{and}
  \bibinfo{author}{\bibfnamefont{J.-M.} \bibnamefont{L'Hermite}},
  \bibinfo{journal}{Phys. Rev. Lett.} \textbf{\bibinfo{volume}{99}},
  \bibinfo{pages}{193401} (\bibinfo{year}{2007}).

\bibitem[{\citenamefont{de~Heer et~al.}(1987)\citenamefont{de~Heer, Knight,
  Chou, and Cohen}}]{nuo_sodium}
\bibinfo{author}{\bibfnamefont{W.~A.} \bibnamefont{de~Heer}},
  \bibinfo{author}{\bibfnamefont{W.~D.} \bibnamefont{Knight}},
  \bibinfo{author}{\bibfnamefont{M.~Y.} \bibnamefont{Chou}}, \bibnamefont{and}
  \bibinfo{author}{\bibfnamefont{M.~L.} \bibnamefont{Cohen}}, in
  \emph{\bibinfo{booktitle}{Solid State Physics}}, edited by
  \bibinfo{editor}{\bibfnamefont{F.}~\bibnamefont{Seitz}} \bibnamefont{and}
  \bibinfo{editor}{\bibfnamefont{D.}~\bibnamefont{Turnbull}}
  (\bibinfo{publisher}{Academic, New York}, \bibinfo{year}{1987}),
  vol.~\bibinfo{volume}{40}, p.~\bibinfo{pages}{93}.

\bibitem[{\citenamefont{Br\'echignac et~al.}(1989)\citenamefont{Br\'echignac,
  Cahuzac, Leygnier, and Weiner}}]{Brechignac1989}
\bibinfo{author}{\bibfnamefont{C.}~\bibnamefont{Br\'echignac}},
  \bibinfo{author}{\bibfnamefont{P.}~\bibnamefont{Cahuzac}},
  \bibinfo{author}{\bibfnamefont{J.}~\bibnamefont{Leygnier}}, \bibnamefont{and}
  \bibinfo{author}{\bibfnamefont{J.}~\bibnamefont{Weiner}},
  \bibinfo{journal}{J. Chem. Phys.} \textbf{\bibinfo{volume}{90}},
  \bibinfo{pages}{1492} (\bibinfo{year}{1989}).

\bibitem[{\citenamefont{Kusche et~al.}(1999)\citenamefont{Kusche, Hippler,
  Schmidt, von Issendorff, and Haberland}}]{Kusche1999}
\bibinfo{author}{\bibfnamefont{R.}~\bibnamefont{Kusche}},
  \bibinfo{author}{\bibfnamefont{T.}~\bibnamefont{Hippler}},
  \bibinfo{author}{\bibfnamefont{M.}~\bibnamefont{Schmidt}},
  \bibinfo{author}{\bibfnamefont{B.}~\bibnamefont{von Issendorff}},
  \bibnamefont{and}
  \bibinfo{author}{\bibfnamefont{H.}~\bibnamefont{Haberland}},
  \bibinfo{journal}{Eur. Phys. J. D} \textbf{\bibinfo{volume}{9}},
  \bibinfo{pages}{1} (\bibinfo{year}{1999}).

\end{thebibliography}



\end{document}